\documentclass{article}
\usepackage{amsmath}
\usepackage{amssymb}
\usepackage{graphicx}

\pdfoutput=1

\def\be{\begin{eqnarray}}
\def\ee{\end{eqnarray}}
\def\nn{\nonumber}

\def\beq{\begin{equation}}
\def\eeq{\end{equation}}

\newdimen\linethick  \linethick=0.4pt
\newdimen\hboxitspace    \hboxitspace=5pt
\newdimen\vboxitspace    \vboxitspace=5pt

\def\fr#1{%
\beq
\vcenter{
\hrule height\linethick
          \hbox{\vrule width\linethick
                \kern\hboxitspace
                \vbox{\kern\vboxitspace
                      \hbox{$\begin{array}{c}\displaystyle#1
         \end{array}$}%
                      \kern\vboxitspace}%
                \kern\hboxitspace
                \vrule width\linethick}%
          \hrule height\linethick}%
\eeq}

\hoffset= - 1.0in         

\title{{\bf Proving AGT conjecture as HS duality:\\
extension to five dimensions } \vspace{.2cm}}
\author{{\bf A.Mironov}\footnote{ {\small {\it
Lebedev Physics Institute} and {\it ITEP, Moscow, Russia}};
mironov@itep.ru; mironov@lpi.ru}, {\bf A.Morozov}\thanks{{\small
{\it ITEP, Moscow, Russia}}; morozov@itep.ru},
{\bf
Sh.Shakirov}\thanks{{\small {\it Department of Mathematics, University
Of California, Berkeley, USA} and {\it ITEP, Moscow, Russia}};
shakirov@math.berkeley.edu; shakirov@itep.ru}, {\bf
A.Smirnov}\thanks{{\small {\it ITEP Moscow, Russia and MIPT,
Dolgoprudny, Russia}}; asmirnov@itep.ru}\date{ }}

\begin{document}
 \maketitle

\vspace{-5.0cm}

\begin{center}
\hfill FIAN/TD-06/11\\
\hfill ITEP/TH-09/11\\
\end{center}

\vspace{3.5cm}

\centerline{ABSTRACT}

\bigskip

{\footnotesize We extend the proof from \cite{Dir},
which interprets the AGT relation as the Hubbard-Stratonovich duality relation
to the case of $5d$ gauge theories. This involves an additional
$q$-deformation. Not surprisingly, the extension turns out to be
trivial: it is enough to substitute all relevant numbers by
$q$-numbers in all the formulas, Dotsenko-Fateev integrals by the
Jackson sums and the Jack polynomials by the MacDonald ones. The problem
with extra poles in individual Nekrasov functions continues to
exist, therefore, such a proof works only for $\beta = 1$, i.e. for
$q=t$ in MacDonald's notation. For $\beta\ne 1$ the conformal blocks are related
in this way to a non-Nekrasov decomposition of the LMNS partition function
into a double sum over Young diagrams.}

\bigskip

\section{Introduction}

The AGT relation \cite{AGT}-\cite{Dir} is a
particular version of the AdS/CFT correspondence
and, more generally, of a gauge/string duality,
which is very interesting, because it is a very concrete
and explicit {\it quantitative} relation between the $2d$ conformal
blocks \cite{CFT} and the instanton partition functions
\cite{LMNS}.
At the same time, it is highly non-trivial, both conceptually
and technically, and a clear proof is still unavailable.
A proof is known in  some simple particular cases \cite{power,2pap},
while in general it is reduced to various technically involved recursion schemes
in \cite{FaLi,poles}, \cite{AL} and \cite{towaproof}.
Recently, in \cite{Dir} we used one of the approaches,
based on the Dotsenko-Fateev-style representation of conformal blocks
\cite{power,DV,Wyl,AGTmamo,ito,MMMS,MManM}
and the character calculus \cite{charex} from matrix model theory,
to cook up a proof based on the standard duality argument.
Namely, one can find a quantity, which involves a double sum,
and two different summation orders provide the two sides of the
AGT relation.
In this particular case this is a sum over characters,
also averaged over time-variables:
if the sum is taken first, one obtains Dotsenko-Fateev integrals
of \cite{MMMS} in the form of \cite{ito};
if the average is taken first, one obtains sum of the Nekrasov functions
\cite{Nek}.
Unfortunately, it works so simple only for $\beta=1$,
otherwise, particular Nekrasov functions have extra poles,
which somehow disappear from the sum and are not seen at
the conformal block side of the AGT relation:
what this really means and how these fictitious poles
should be interpreted and handled within the AGT context,
remains a mystery.

Instead for $\beta\ne 1$ the Hubbard-Stratonovich duality provides another, non-Nekrasov
decomposition of the LMNS partition function \cite{LMNS} into a double sum over Young
diagrams, which may have its own significance (one natural way to proceed in this direction
is to extend the results of \cite{Dir} from the spherical 4-point to the arbitrary
conformal block).
In this letter we consider a natural $q$-deformation
of \cite{Dir}, which corresponds to the straightforward
generalization of Seiberg-Witten theory \cite{SW,SWint}, of Nekrasov calculus
and of the AGT relation from $4d$ to $5d$ theories.
Such an extension has already been addressed in the literature:
in \cite{5dSW,5dB} and \cite{Awata,Awata2,Yan}.
It is well-known to be straightforward and should not bring
any surprises.
At the same time, it involves some technicalities in
character calculus, because it involves
the MacDonald polynomials in the role of characters and
the Jackson sums in the role of open-contour integrals.
As usual, $q$-deformation is the level, where all technical features
look most natural and all formulas become most transparent.
Also it is a natural step towards further generalization:
to somewhat more general Kerov polynomials
and to $6d$ theories, the very interesting in the AGT context.
The last, but not least:
the $5d$ deformation seems to play a role
in "3d" extensions of the AGT relation \cite{Kanno,MMSm3dAGT},
which are supposed to involve $3d$ Chern-Simons
theory \cite{CS} and knot invariants \cite{Gukov,Jap}.

As expected, since all the formulas of \cite{Dir}
for the $N_f=2N_c=4$ are nicely factorizable,
they are directly generalized to $q\neq 1$,
by substitution of all the factors by their $q$-number
counterparts:
\be
n\rightarrow [n]_q=\dfrac{\ \ 1-q^n}{1-q}
\ee
We do not consider here the "pure gauge limit" part of the
story: it is again straightforward, but the proper
$q$-version of the Brezin-Gross-Witten unitary $\beta$-ensemble \cite{PGL}
deserves separate consideration.

\section{Four dimensions}

We start with outlining the main aspects of the proof of the standard AGT conjecture
in four dimensional case for $\beta=1$. In $SU(2)$ case the AGT conjecture claims
that the instanton part of the four-dimensional ${\cal{N}}=2$ superconformal field
theory coincides with the $4$-point conformal block in $2d$
CFT\footnote{Here $B(\Delta_i, \Delta, c|\,\Lambda)$ is the 4-point conformal block
with fields located at $0$, $\Lambda$, $1$ and  $\infty$.
We use $\Lambda$ to denote the double ratio of
four coordinates instead of the more conventional
$q$ or $x$, because these letters are used for
other purposes in the present text.
Physically, $\Lambda = e^{2\pi i \tau}$,
where $\tau$ is the bare coupling constant,
it turns into dimensional $\Lambda_{QCD}$
after dimensional transmutation when some
of the masses $m_1,\ldots,m_4$ tend to infinity. }:
\be
Z_{Nek}^{4d}(\epsilon_i, \mu_i,a|\,\Lambda)=B(\Delta_i, \Delta, c|\,\Lambda)
\ee
under certain identification of the parameters $\{\epsilon_i, \mu_i,a\}$ and
$\{\Delta_i, \Delta, c\}$. The Nekrasov partition function has the form of double
expansion over two sets of Young diagrams:
\be
\label{NP}
Z_{Nek}^{4d}(\Lambda)=\sum\limits_{A,B} \,N_{A,B} (\epsilon_i, \mu_i,a)\, \Lambda^{|A|+|B|}
\ee
where the coefficients $N_{A,B}$ are the Nekrasov functions corresponding to the Young
diagrams $A$ and $B$.

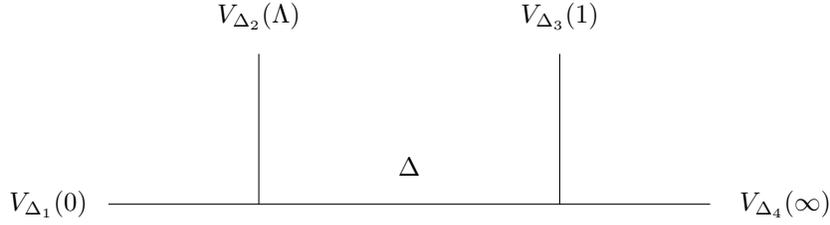
\begin{figure}\begin{center}
\unitlength 1mm 
\linethickness{0.4pt}
\ifx\plotpoint\undefined\newsavebox{\plotpoint}\fi 
\begin{picture}(100,20)(0,8)
%
%
\put(10,10){\line(1,0){80}}
\put(30,10){\line(0,1){20}}
\put(70,10){\line(0,1){20}}
\put(2,10){\makebox(0,0)[cc]{$V_{\Delta_1}(0)$}}
\put(30,35){\makebox(0,0)[cc]{$V_{\Delta_2}(\Lambda)$}}
\put(70,35){\makebox(0,0)[cc]{$V_{\Delta_3}(1)$}}
\put(100,10){\makebox(0,0)[cc]{$V_{\Delta_4}(\infty)$}}
\put(50,15){\makebox(0,0)[cc]{$\Delta$}}
\end{picture}
\caption{\footnotesize{
Feynman diagram for the 4-point conformal block.
}}\label{4block}
\end{center}\end{figure}

It is well-known that the $\Lambda$-expansion of the conformal block based on the
operator product expansion (OPE) has the form of the sum over two Young diagrams. This
OPE procedure is extensively reviewed in the CFT literature
\cite{CFT,AGTmmm,MMMM,nonconf};  in the particular
$4$-point case shown in the Fig.\ref{4block}, it gives:
\be
\label{opem}
B(\Delta_{1},\Delta_{\Delta_2}, \Delta_{3},\Delta_{\Delta_4},\Delta,c\,|\,\Lambda)=
\sum\limits_{A,B}\, \Lambda^{\frac{1}{2}(|A|+|B|)}\,
\gamma_{\Delta_1\Delta_2\Delta;A} Q_{\Delta}^{-1}( A,B )
\gamma_{\Delta \Delta_3\Delta_4;B}
\ee

\noindent
where $\gamma_{\Delta_1\Delta_2\Delta_3;A}$ are the structure coefficients of the OPE
algebra, and $Q$ is the Shapovalov form of the Virasoro algebra:
\be
Q_{\Delta}(A,B)=< \Delta| L_{A} L_{-B} |\Delta >
\ee
$\gamma_{\Delta_1\Delta_2\Delta_3;A}$ are known explicitly, while $Q_{\Delta}(A,B)$
can be calculated level by level (see, e.g., \cite{AGTmmm}) and one can directly
construct the $\Lambda$-expansion. However, this expansion \textit{does not
coincide (!)} with the double expansion of the Nekrasov partition function (\ref{NP}).
Indeed, the Shapovalov form  $Q_{\Delta}(A,B)$ is not zero only for descendants of
the same level, which means that only the Young diagrams with $|A|=|B|$ contribute
to the sum (\ref{opem}), but there is no such a restriction in (\ref{NP}).

The appropriate double expansion of the $4$-point conformal block comes
 from the free field representation of correlator. As was shown in
 \cite{ito,MManM,Dir},
utilizing the Dotsenko-Fateev integral representation \cite{DF}, the conformal block
can be represented as a double average over the two independent Selberg ensembles:
\be
\label{cb}
B(\Delta_i, \Delta, c|\,\Lambda)=\left< \left<  \prod\limits_{i=1}^{N_+}(1-\Lambda x_{i})^{v_-} \prod\limits_{j=1}^{N_-}(1-\Lambda y_{j})^{v_+}  \prod\limits_{i=1}^{N_+} \prod\limits_{j=1}^{N_-} (1-\Lambda x_i y_j)^{2\beta } \right>_+\right>_-
\ee
Here the average goes over two ensembles (labeled by symbols $+$ and $-$)
of variables $x_1,...x_{N_{+}}$ and $y_{1},..., y_{N_{-}}$ ("eigenvalues of matrix
models"):

$$
\Big<  f\big(x_1, \ldots, x_{N_{+}}\big)  \Big>_{+} = \dfrac{1}{Z_+}\
\int\limits_{0}^{1} dx_1 \ldots \int\limits_{0}^{1} dx_{N_{+}} \prod\limits_{i<j} (x_i - x_j)^{2 \beta} \prod\limits_{i} x_i^{u_+} (x_i - 1)^{v_+} \ f\big(x_1, \ldots, x_{N_{+}}\big)
$$
$$
\Big<  f\big(y_1, \ldots, y_{N_{-}}\big)  \Big>_{-} = \dfrac{1}{Z_-}\
\int\limits_{0}^{1} dy_1 \ldots \int\limits_{0}^{1} dy_{N_{-}} \prod\limits_{i<j} (y_i - y_j)^{2 \beta} \prod\limits_{i} y_i^{u_{-}} (y_i - 1)^{v_{-}} \ f\big(y_1, \ldots, y_{N_{-}}\big)
$$
with the normalization constants
$$
Z_{\pm} =
\int\limits_{0}^{1} dz_1 \ldots \int\limits_{0}^{1} dz_{N_{\pm}} \prod\limits_{i<j} (z_i - z_j)^{2 \beta} \prod\limits_{i} z_i^{u_{\pm}} (z_i - 1)^{v_{\pm}}
$$
This matrix model representation of the conformal block is very convenient for
analysis of its $\Lambda$-expansion, moreover, utilizing the standard matrix model
technique of character expansion for each set of variables one can rewrite (\ref{cb})
as a double expansion over two sets of Young diagrams. Indeed, let us denote by
 $I$ the function which is averaged in (\ref{cb}), then one has:
$$
I=\prod\limits_{i=1}^{N_+}(1-q x_{i})^{v_-} \prod\limits_{j=1}^{N_-}(1-q y_{j})^{v_+}  \prod\limits_{i=1}^{N_+} \prod\limits_{j=1}^{N_-} (1-q x_i y_j)^{2\beta }= \ \ \ \ \ \ \ \ \ \ \ \ \ \ \ \ \ \
$$
$$
 \ \ \ \ \ \  \ \ \ =\exp\Big( v_-\sum\limits_{i=1}^{N_+}\ln(1-\Lambda x_{i})+v_+\sum\limits_{j=1}^{N_-}\ln(1-\Lambda y_{i}) +2 \beta \sum\limits_{i=1}^{N_+}\sum\limits_{j=1}^{N_-} \ln(1-\Lambda x_{i} y_{j}) \Big)
$$

\be
\label{pp}
=\exp\Big( -\sum_{k=1}^{\infty} \dfrac{\Lambda^k}{k} p_{k} v_{-}-\sum_{k=1}^{\infty} \dfrac{\Lambda^k}{k} \widetilde  p_{k} v_{+} -2 \beta \sum\limits_{k=1}^{\infty} \dfrac{\Lambda^k}{k} p_{k}\widetilde  p_{k}  \Big)\ \ \ \ \ \ \ \ \ \ \ee
where in the last step we expanded the logarithms into the powers of $\Lambda$ and
denoted
\be
p_{k}=\sum\limits_{i=1}^{N_+} x_{i}^{k}, \ \ \ {\widetilde  p_{k}}=\sum\limits_{j=1}^{N_-} y_{j}^{k}, \ \ \ \textrm{such that} \ \ \ \sum\limits_{i=1}^{N_+} \ln(1-\Lambda x_{i})=-\sum\limits_{i=1}^{N_+}\sum\limits_{k=1}^{\infty} \dfrac{\Lambda^{k} x_{i}^{k}}{k}=-\sum\limits_{k=1}^{\infty}\dfrac{\Lambda^{k} }{k} p_{k}
\ee
We rewrite (\ref{pp}) in the form \cite{ito,MManM}
\be
\label{pp2}
I=\exp\left( \beta \sum\limits_{k=1}^{\infty} \dfrac{\Lambda^k}{k} p_{k}\Big(-{\widetilde  p_{k}}-\dfrac{v_{-}}{\beta}  \Big) \right)\,\exp\left( \beta \sum\limits_{k=1}^{\infty} \dfrac{\Lambda^k}{k} {\widetilde  p_{k}}\Big(-p_{k}-\dfrac{v_{+}}{\beta}  \Big) \right)
\ee
The final step that one needs in order to expand (\ref{pp}) into the sum of characters
is the Cauchy completeness formula for the Jack polynomials:
\be
 \exp\Big(\beta \sum\limits_{k=1}^{\infty} \, \frac{p_{k} {\widetilde p}_{k}}{k}\Big)=\sum\limits_{R} \, j_{R}(p_{k}) j_{R}({\widetilde p}_{k})
\ee
where $j_{R}$ is the normalized Jack polynomial (with deformation parameter $\beta$)
corresponding to the representation $R$, and the sum runs over all representations of
$GL(\infty)$ (over all the Young diagrams $R$). Utilizing this formula for (\ref{pp2})
one finally finds
\be
I=\sum\limits_{A,B} \Lambda^{|A|+|B|}   j_{B}(p_{k}) j_{B}\Big( -{\widetilde p}_{k}-\dfrac{v_{-}}{\beta}\Big) j_{A} ({\widetilde p}_k ) j_{A} \Big( -p_k-\dfrac{v_{+}}{\beta} \Big)
\ee
Note that, due to presence of the term $2 \beta p_{k} {\widetilde p}_k  $ in (\ref{pp}),
the expansion goes over a set of two Young diagrams $A$ and $B$. We find that the
$\Lambda$-expansion of the conformal block takes the form similar to the expansion of
the Nekrasov partition function:
\fr{
\label{cec}
B(\Lambda)=\sum_kB_k\Lambda^k=\sum\limits_{A,B} \Lambda^{|A|+|B|}   \Big<j_{A} ( -p_k-v_+ ) j_{B}(p_{k})\Big>_{+} \Big<j_{A} ({\widetilde p}_k )  j_{B}( -{\widetilde p}_{k}-v_{-})\Big>_{-}
}
Comparing both sides of (\ref{NP}) and (\ref{cec}), the AGT conjecture states that
\be
\label{bs}
\sum _{A,B} N_{A,B} = \sum _{A,B} \Big<j_{A} ( -p_k-{v_+\over\beta} ) j_{B}(p_{k})\Big>_{+}
\Big<j_{A} ({\widetilde p}_k )  j_{B}( -{\widetilde p}_{k}-{v_{-}\over\beta})\Big>_{-}
\ee
But really exciting is that the identity becomes termwise in the case of $\beta=1$
(corresponding to the case of $\epsilon_1+\epsilon_2=0$ on the side of the
Nekrasov function) \cite{Dir}:
\be
\label{twr}
\left.N_{A,B}\right|_{\epsilon_{1}+\epsilon_{2}=0} = \left.\Big<j_{A} ( -p_k-{v_+\over\beta} )
j_{B}(p_{k})\Big>_{+} \Big<j_{A} ({\widetilde p}_k ) j_{B}( -{\widetilde p}_{k}-{v_{-}\over\beta})\Big>_{-}\right|_{\beta=1}
\ee
In this way, the AGT relation is interpreted as a standard duality of the
Hubbard-Stratonovich type, see Fig.\ref{dual}:
\be\label{HS}
\sum_{a,b}\left(\sum_i X_i^a X_i^b\right)\left(\sum_j X_j^a X_j^b\right)
=\sum_{a,b,i,j} X_i^a X_i^b X_j^a X_j^b
= \sum_{i,j}\left(\sum_a X_i^a X_j^a\right)\left(\sum_b X_i^b X_j^b\right)
\ee
In our case the role of $X_{i}^{a}$ is played by the symmetric polynomials
$j_{A}(p_{k})$, summation over $a,b$ corresponds to the summation over the
Young diagrams and summation over $i$ and $j$ is the averaging over two independent
ensembles.
\begin{figure}\label{dual}
\unitlength 0.5mm 
\linethickness{0.4pt}
\ifx\plotpoint\undefined\newsavebox{\plotpoint}\fi 
\begin{picture}(00,173.5)(-15,-20)
\put(129.5,129.25){\line(1,1){19}}
\put(148.5,148.5){\line(-1,1){22}}
\put(153.5,148.5){\line(1,1){22}}
\put(153.5,148.5){\line(1,-1){19}}
\multiput(150,151)(.664063,.664063){33}{{\rule{.4pt}{.4pt}}}
\multiput(150,151)(-.647059,.647059){35}{{\rule{.4pt}{.4pt}}}
\multiput(151.18,145.68)(-.65179,-.65179){29}{{\rule{.4pt}{.4pt}}}
\multiput(151.18,145.68)(.64286,-.66964){29}{{\rule{.4pt}{.4pt}}}
\put(124.25,118){\vector(-2,-3){15}}
\put(30,89.75){\line(1,-1){19.25}}
\put(49,70.5){\line(-1,-1){17.5}}
\multiput(32.68,92.43)(.66667,-.65){31}{{\rule{.4pt}{.4pt}}}
\multiput(52.68,72.93)(.98913,0){47}{{\rule{.4pt}{.4pt}}}
\multiput(35.18,50.93)(.65385,.67308){27}{{\rule{.4pt}{.4pt}}}
\multiput(52.18,68.43)(.984043,-.005319){48}{{\rule{.4pt}{.4pt}}}
\put(101.75,70.75){\line(1,1){20}}
\put(102,70.5){\line(1,-1){19}}
\multiput(98.43,67.93)(.65179,-.67857){29}{{\rule{.4pt}{.4pt}}}
\multiput(98.68,73.18)(.65,.66667){31}{{\rule{.4pt}{.4pt}}}
\put(199.25,93.5){\line(1,-1){17}}
\put(216.25,76.5){\line(0,-1){39}}
\put(216.25,37.5){\line(-1,-1){17.75}}
\put(220.5,76.5){\line(1,1){19.5}}
\put(220.5,76.25){\line(0,-1){39.25}}
\put(220.5,37){\line(1,-1){17.5}}
\multiput(202.43,96.93)(.65,-.66){26}{{\rule{.4pt}{.4pt}}}
\multiput(218.68,80.43)(.64286,.65179){29}{{\rule{.4pt}{.4pt}}}
\multiput(201.68,17.93)(.65,.65){26}{{\rule{.4pt}{.4pt}}}
\multiput(217.93,34.18)(.67308,-.65385){27}{{\rule{.4pt}{.4pt}}}
\put(178.5,122){\vector(1,-1){15}}
\put(0,-10){\makebox(0,0)[cc]{$\sum_{A,B}N_{AB}$}}
\put(23,-10){\makebox(0,0)[cc]{=}}
\put(93,-10){\makebox(0,0)[cc]
{\hspace{-6ex}$\overbrace{\sum_{AB}\int_{x}  j_A(x)j_{B}(x)\int_y j_A(y) j_B(y)}$\hspace{-2ex}}}
\put(152,-10){\makebox(0,0)[cc]{\hspace{-2ex}=}}
\put(233,-10){\makebox(0,0)[cc]
{\hspace{-12ex}$\overbrace{\int_{x,y} \sum_A j_A(x)j_A(y)\sum_B  j_B(x)j_B(y)}$\hspace{-2ex}}}
\put(285,-10){\makebox(0,0)[cc]{=}}
\put(297,-10){\makebox(0,0)[cc]{$B(\Lambda)$}}
\put(74,38.75){\vector(0,-1){15}}
\put(216,18.75){\vector(0,-1){15}}
\end{picture}
\caption{Picture of the Nekrasov functions/conformal block duality expressed by the
Hubbard-Stratonovich type formula (\ref{HS}). The symbol $\int_z$ here denotes
integration with the Selberg measure over variables $z_i$, and the symbol
$\sum_A$ denotes summation over all Young diagrams $A$.}
\end{figure}
Unfortunately, relation (\ref{twr}) is broken at $\beta \neq 1$ (relation
(\ref{bs}), of course, remains true in this case as well). In this case the individual
Nekrasov function has more poles then the whole sum (\ref{bs}). These extra poles
puzzle \cite{Dir} remains unresolved and the interpretation of the original AGT conjecture as
a Hubbard-Stratonovich duality is still missed in the case of $\beta \neq 1$.
Instead, (\ref{cec}) provides an alternative (modified) AGT conjecture which is, perhaps,
even more interesting and useful than the original one.
The items of the bi-Selberg decomposition (\ref{cec})
have no extra poles, but the numerators do
{\it not} factorize into linear factors,
as in the Nekrasov decomposition.
The example of the first level $|A|+|B|=1$ is
already fully representative:
\be
B_1 = \frac{(a+m_1)(a+m_2)(a+m_3)(a+m_4)}{2a(2a+\epsilon)}
+  \frac{(a-m_1)(a-m_2)(a-m_3)(a-m_4)}{2a(2a-\epsilon)}
= \nn \\
=  \frac{\Big((a+m_1)(a+m_2)-\epsilon(m_1+m_2)\Big)
(a+m_3)(a+m_4)}{(4a^2-\epsilon^2)} +
\frac{(a-m_1)(a-m_2)\Big((a-m_3)(a-m_4)-\epsilon(m_3+m_4)\Big)
}{(4a^2-\epsilon^2)}
\label{twodeco}
\ee
where the first line is the Nekrasov decomposition,
while  the second line is the bi-Selberg one in (\ref{cec}).
Clearly, the two decompositions are different, but
coincide for $\epsilon=\epsilon_1+\epsilon_2=0$, i.e.
for $\beta=1$. In fact, in addition to (\ref{cec}), there is also an alternative decomposition:
$$
B(\Lambda)=\sum\limits_{A,B} \Lambda^{|A|+|B|}   \Big<j_{A} (p_k+{v_+\over\beta} )
j_{B}(p_{k})\Big>_{+}
\Big<j_{A} (-{\widetilde p}_k )  j_{B}( -{\widetilde p}_{k}-{v_{-}\over\beta})\Big>_{-}
$$

However, at level 1 it is indistinguishable from (\ref{cec}) and we do not add the extra
line to (\ref{twodeco}). Note that no one of the three correlators:
$\Big<j_{A} (p_k+v/\beta ) j_{B}(p_{k})\Big>$, $\Big<j_{A} (-p_k-v/\beta ) j_{B}(p_{k})\Big>$,
$\Big<j_{A} (-p_k-v/\beta ) j_{B}(-p_{k})\Big>$ is factorizable at $\beta\ne 1$. The only
factorizable correlator is $\Big<j_{A} (p_k+w ) j_{B}(p_{k})\Big>$, however,
$w\ne v/\beta$ for $\beta\ne 1$ (see (\ref{shift}) below).

Leaving this problem, the generalization of (\ref{twr}) to the five-dimensional case is
straightforward. As was noted in \cite{5dSW,5dB} every $4d$ Seiberg-Witten theory
can be generalized to the $5d$ case by an appropriate $q$-deformation,
with the deformation parameter $q=e^{-\hbar R} $, with $R$ being
radius of the compact fifth dimension, so that in the case of $R=0$ or $q=1$ one
returns to the standard four-dimensional theory. In particular, the
deformation of the four-dimensional Nekrasov function to five
dimensions is very simple: all the factors of the four-dimensional
Nekrasov function are substituted by their
$q$-number counterparts
\be
n\rightarrow [n]_q=\dfrac{\ \ 1-q^n}{1-q}
\ee
The aim of this paper is to describe the appropriate $q$-deformation of relation
(\ref{twr}). Some progress in this direction has been already made in \cite{Awata2}
where the $q$-deformed conformal block is fixed by the $q$-Virasoro algebra.
The free field representation for the $q$-deformed vertex operators can be found in
\cite{Awata}.

Here we do not consider all the preliminary steps, and start directly from
$q$-deformation of the double average (\ref{cb}). Such a $q$-deformation can be
straightforwardly written using the usual properties of $q$-deformation. All one needs,
is to change the factors and integrals in (\ref{cb}) by their $q$-counterparts, the
rules are as follows
\begin{itemize}
\item all power-like factors in (\ref{cb}) are substituted with the products:
\be
\label{rb}
(1-x)^{a}\rightarrow \prod\limits_{k=0}^{a-1} (1-q^k x)
\ee
\item the Van-der-Monde determinant (the Jack measure) is replaced by the
MacDonald measure:
\be
\prod_{1\leq i<j \leq N} (x_{i}-x_{j})^{2 \beta}\rightarrow \Delta^{MC} (x) \rightarrow \prod_{i \neq j} \prod_{k=0}^{\beta-1} (x_i- q^k x_{j})
\ee
\item The integrals in the Selberg  average are replaced by the $q$-Jackson
integrals (see (\ref{ji}) in the Appendix for the definition):
\be
\label{re}
\int\limits_{0}^{1} dz \rightarrow  \int\limits_{0}^{1} d_{q} z
\ee
\end{itemize}
In complete analogy with the four-dimensional case, these simple rules lead to the
Jackson integral representation of the five-dimensional conformal block and, further,
the Nekrasov functions. Similar to the four-dimensional case, formula (\ref{twr})
works only at $\beta=1$, and the problem of extra poles of the Nekrasov functions
remains unresolved.

As a by product of this research, we found a nice, completely factorized formula for
the average of two MacDonald polynomials (\ref{tm}). Similar to the Nekrasov functions,
this average is completely factorized into linear multiples, but gives the
Nekrasov function only at $\beta=1$.

\section{AGT in five dimensions}
\subsection{Nekrasov Functions \label{nfs}}

The instanton part of the five-dimensional $SU(N)$ partition function with $N_f=2N$
fundamentals has form of the sum over $N$ Young diagrams $Y_{i}, (i=1...N)$:
\be
Z^{5d}_{Nek}(\Lambda)=\sum\limits_{Y_{1},...,Y_{N}} \, N_{Y_1,...,Y_{N}}\,
\tilde\Lambda^{|Y_{1}|+...+|Y_{N}|}
\ee
and the coefficients of expansion are \cite{Awata}
\be
\label{NF}
N_{Y_1,...,Y_{N}}= \Big( v^{-N} \,\prod\limits_{j=1}^{N}
(Q_{j}^{+})^{\frac{1}{2}}(Q_{j}^{-})^{-\frac{1}{2}} \,\Big)^{|Y_{1}|+...+|Y_{N}|}
\prod\limits_{i,j=1}^{N}\,\dfrac{{\cal{N}}_{Y_{i},[]} (v Q_{i}/Q_{j}^{+}) {\cal{N}}_{[],Y_{i}} (v Q_{j}^{-}/Q_{i})   }{ {\cal{N}}_{Y_i, Y_{j}} ( Q_{i}/Q_{j}) }
\ee
with
\be
\label{F}
{\cal{N}}_{A,B}(Q)=\prod\limits_{(i,j)\in A} \Big( 1-Q q^{\textrm{Leg}_{A}(i,j)} t^{\textrm{Arm}_{B}(i,j)+1} \Big)
\prod\limits_{(i,j)\in B}\Big( 1-Q q^{-\textrm{Leg}_{B}(i,j)-1} t^{-\textrm{Arm}_{A}(i,j)}  \Big)
\ee
where $ v=(q/t)^{1/2}$ and $[]$ denotes the empty Young diagram.
The first multiplier in (\ref{NF}) can be put unit by rescaling the expansion
parameter $\Lambda$, we keep it in order to make the Nekrasov functions (\ref{NF})
symmetric in masses.

\noindent

\raisebox{-24mm}{ \includegraphics[height=50mm]{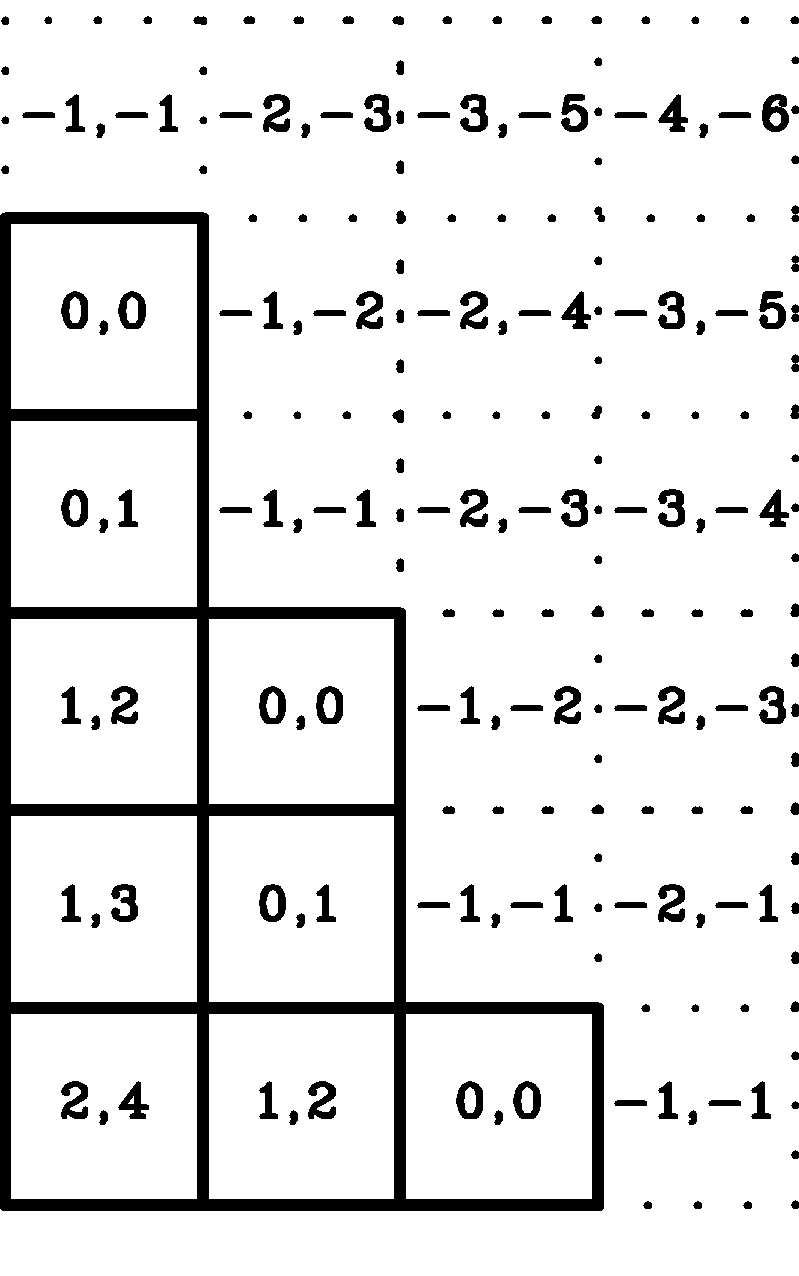}
\put(-50,-3){ Fig.3  } }
\parbox{13cm}{

\noindent
The parameters $t$ and $q$ are related with the $\Omega$-background parameters as
$q=e^{R \epsilon_{2}}$ and $t=e^{-R \epsilon_{1}}$, where $R$ is the radius of the
compact fifth dimension. The remaining parameters in (\ref{NF}) are related with the
v.e.v. of scalar fields $a_{i}$ and the masses of fundamentals $m_i=\mu_{i}
\sqrt{\epsilon_1\epsilon_2}$ as
follows:
\be
Q_{i}=q^{a_{i}},\ \ \ Q_{i}^{+}=q^{-\mu_{i}}, \ \ \ Q_{i}^{-}=q^{-\mu_{N+i}}
\ee
Note that in \cite{Dir} we used different normalization for the
v.e.v.'s $a_{i}$ and the masses $\mu_{i}$:
$$
a_{i}\rightarrow \epsilon_{2} a_{i},\ \ \ \mu_{i} \rightarrow \epsilon_{2} \mu_{i}
$$
For the arbitrary Young diagram $Y$, the symbols ${\rm Arm}_Y(i,j)$
 and ${\rm Leg}_Y(i,j)$ denote
 the arm-length and leg-length of the box $(i,j)$ in the Young diagram $Y$
respectively. Algebraically,
these lengths are given by the expressions
\begin{align}
{\rm Arm}_Y(i,j) = Y^{\prime}_j - i, \ \ \ {\rm Leg}_Y(i,j) = Y_i - j
\end{align}}

\noindent
where $Y^{\prime}$ stands for the transposed Young diagram. Note that functions
$\textrm{Arm}_{Y}(i,j)$ and $\textrm{Leg}_{Y}(i,j)$ can take negative values for
$(i,j)$ outside the Young diagram $Y$.
In Fig.3 we give an example of the Young diagram $Y=[5,3,1]$ with the
corresponding lengthes $( \textrm{Leg}_{Y}(i,j),\textrm{Arm}_{Y}(i,j) )$ both
within the diagram $Y$ and outside it.

In the case of $N=2$, the partition function takes the form
\be
\label{Nd}
Z^{5d}_{Nek}(\Lambda)=\sum\limits_{A,B} N_{A,B} \tilde\Lambda^{|A|+|B|}
\ee
and the coefficients can be rewritten in the form used in \cite{Dir}:
\be
\label{Shf}
N_{A,B}=\dfrac{\prod\limits_{k=1}^{2} f^{+}_{A}(\mu_k+a)f^{+}_{B}(\mu_k-a)\prod\limits_{k=3}^{4} f^{-}_{A}(\mu_k+a)f^{-}_{B}(\mu_k-a)
}{g_{A,A}(0) g_{A,B}(-2a) g_{B,A} (2 a) g_{B,B} (0)}\, q^{\frac{-\mu_1-\mu_2+\mu_3+\mu_4-2(1-\beta)}{2}( |A|+|B|)}
\ee
such that all the functions are some products of $q$-numbers:
\be
\label{f}
f_{A}^{\pm}(x)=\prod\limits_{(i,j)\in A} [\pm x \mp i \beta \pm j \mp \frac{1}{2}(1-\beta)]_{q},
\ee
and
\be
\label{g}
g_{A,B}(x)=\prod\limits_{(i,j)\in A } [x+\beta \textrm{Arm}_{A}(i,j)+\textrm{Leg}_{B}(i,j)+\beta]_{q}[-x-\beta \textrm{Arm}_{A}(i,j)-\textrm{Leg}_{B}(i,j)-1]_{q}
\ee
where we used the following definition of $\beta$:
\be
t=q^{\beta},\ \ \ \beta=-\dfrac{\epsilon_1}{\epsilon_2}
\ee
As we shall see in the case of $N=2$ $\tilde\Lambda$ is actually slightly different
from the $\Lambda$-parameter of the conformal block, that is,
\fr{
\tilde\Lambda=\Lambda q^{\gamma}}
with
\be
\gamma=\sum\limits_{k=1}^{4} \frac{\mu_{k}}{2}+1
\ee

\subsection{ Dotsenko-Fateev integral }

The Dotsenko-Fateev integral representation for the $5d$ conformal block is an
appropriate $q$-deformation of the four-dimensional double average (\ref{cb}).
Similar to four dimensions, this representation can be constructed by utilizing the
free field representation of the conformal block, the corresponding
$q$-deformed vertex operators being described in \cite{Awata}. In fact,
the $q$-deformations of all factors in (\ref{cb}) are well-known, and, hence,
the proper $q$-version of (\ref{cb}) can be obtained directly by the usual rules
(\ref{rb})-(\ref{re}). In this way, one easily finds\footnote{
Hereafter, for the sake of simplicity we write all the formulas for integer values of
parameters $v_+$, $v_-$ and $\beta$. Note, however, that the extension to
non-integer quantities is very straightforward, see (\ref{nib}) and (\ref{niv}) in
the Appendix. }:
\begin{align}
B^{5d}(\Lambda) = \left< \ \left< \ \ \prod\limits_{i = 1}^{N_+} \prod\limits_{k=0}^{v_{-}-1} (1 - q^k\Lambda\, x_i) \prod\limits_{j = 1}^{N_-} \prod\limits_{k=0}^{v_{
+}-1} (1 - q^k \Lambda\, y_j) \prod\limits_{i = 1}^{N_+} \prod\limits_{j = 1}^{N_-}\prod\limits_{k=0}^{\beta-1} (1 - q^k\Lambda\, x_i y_j)^{2} \ \right>_+ \ \right>_-
\label{NekMamo}
\end{align}
\smallskip\\
The averages are taken over two independent sets
(labeled by symbols $+$ and $-$) of variables $x_1,...,x_{N_{+}}$ and
$y_1,...,y_{N_{-}}$ ("eigenvalues in the matrix model terms") as follows:
\begin{align}
\Big< f \Big>_+ \ = \ \dfrac{1}{S_{+}}
\int\limits_{0}^{1} d_qx_1 \ldots \int\limits_{0}^{1} d_qx_{N_+} \prod\limits_{i\neq j} \prod\limits_{k=0}^{\beta-1} (x_i - q^k\,x_j) \prod\limits_{i} x_i^{u_+} \prod\limits_{k=0}^{v_{+}-1}(1-q^k \,x_i) \ f\big(x_1, \ldots, x_{N_+}\big)
\end{align}
\begin{align}
\Big< f \Big>_- \ = \ \dfrac{1}{S_{-}}
\int\limits_{0}^{1} d_qy_1 \ldots \int\limits_{0}^{1} d_qy_{N_-} \prod\limits_{i\neq j} \prod\limits_{k=0}^{\beta-1} (y_i - q^k\,y_j) \prod\limits_{i} y_i^{u_-} \prod\limits_{k=0}^{v_{-}-1}(1-q^k \,y_i) \ f\big(y_1, \ldots, y_{N_-}\big)
\label{yAverage}
\end{align}
with the normalization constants
\begin{align}
S_{\pm} = \int\limits_{0}^{1} d_qz_1 \ldots \int_0^1d_qz_{N} \prod\limits_{i\neq j} \prod\limits_{k=0}^{\beta-1} (z_i - q^k\,z_j) \prod\limits_{i} z_i^{u_\pm}\prod\limits_{k=0}^{v_{+}-1} (1-q^k\,z_i)
\end{align}
which guarantee $\Big< 1 \Big>_+ = \Big< 1 \Big>_- = 1$.
We show in
section \ref{Pr} that the $q$-deformed $\beta$-ensemble (\ref{NekMamo}), indeed,
correctly reproduces the $5d$ Nekrasov partition function.

\subsection{The AGT conjecture}

As we show in the next subsection, in the case of $N=2$ there is a simple identity
between the five-dimensional conformal block (\ref{NekMamo}) and the five-dimensional
partition function (\ref{Nd}):
\begin{align}
B^{5d}(\Lambda) \ = \ Z_{Nek}^{5d}(\Lambda)
\label{AGTmasses}
\end{align}
with the following identification of parameters:
\be
\label{agt}
& \ N_+ = \dfrac{\epsilon_2}{\epsilon_1}(a - \mu_2)=\displaystyle{{\mu_2-a\over\beta}},
\ N_- = -\dfrac{\epsilon_2}{\epsilon_1}( a + \mu_4)=
\displaystyle{{a + \mu_4\over\beta}}
 \nonumber
 \\ \nonumber
&  \\
& u_+ = \mu_1 - \mu_2 -1-\dfrac{  \epsilon_1}{\epsilon_2}=
\mu_1 - \mu_2 -1+\beta, \ u_- = \mu_3 - \mu_4 -1- \dfrac{  \epsilon_1 }{\epsilon_2} =
\mu_3 - \mu_4 -1+\beta\label{mAGT2}\\
& \nonumber \\
& v_+ = -\mu_1-\mu_2, \ v_- = -\mu_3-\mu_4 \nonumber
\ee
Note that this AGT-identification does not depend on $q$.

\subsection{Bi-Selberg expansion of the conformal block\label{Pr}}

The proof of the AGT conjecture for $\beta=1$ is much similar to the $4d$ case outlined in the
Introduction, where
the proof was based on the expansion of the Dotsenko-Fateev integrand
into the Jack polynomials. Obviously, in the $5d$ case the expansion should be into
the MacDonald polynomials, which are the appropriate $q$-deformation of the Jack
functions.
Denote by $I$ the integrand of (\ref{NekMamo}), then:
$$
I=\prod\limits_{i = 1}^{N_+} \prod\limits_{k=0}^{v_{-}-1}
 (1 - q^k\Lambda\, x_i) \prod\limits_{j = 1}^{N_-} \prod\limits_{k=0}^{v_{
+}-1} (1 - q^k \Lambda\, y_j) \prod\limits_{i = 1}^{N_+} \prod\limits_{j = 1}^{N_-}\prod\limits_{k=0}^{\beta-1} (1 - q^k\Lambda\, x_i y_j)^{2}= \ \ \ \ \ \ \ \ \ \ \ \ \ \ \ \ \ \ \ \ \ \ \ \ \ \ \ \ \ \ \ \ \ \
$$
$$
=\exp\left( \sum\limits_{i=1}^{N_+}\sum\limits_{k=0}^{v_--1} \ln(1-q^k \Lambda x_{i})+
\sum\limits_{j=1}^{N_-}\sum\limits_{k=0}^{v_+-1} \ln(1-q^k \Lambda y_{j})+ 2\sum\limits_{i=1}^{N_+}\sum\limits_{j=1}^{N_-}\sum\limits_{k=0}^{\beta-1} \ln(1-q^k \Lambda x_{i} y_{j})\right)= \ \ \ \ \ \ \ \ \ \
$$
$$
=\exp\left( -\sum\limits_{i=1}^{N_+}\sum\limits_{k=0}^{v_--1} \sum\limits_{m=1}^{\infty} \dfrac{q^{km} {\Lambda}^m x_{i}^m}{m}-
\sum\limits_{j=1}^{N_-}\sum\limits_{k=0}^{v_+-1}\sum\limits_{m=1}^{\infty} \dfrac{q^{km} {\Lambda}^m y_{j}^m}{m} -2\sum\limits_{i=1}^{N_+}\sum\limits_{j=1}^{N_-}\sum\limits_{k=0}^{\beta-1}\sum\limits_{m=1}^{\infty} \dfrac{q^{km} {\Lambda}^m x_{i}^m y_{j}^m}{m} \right)=
$$
\begin{align}
=\exp\left( -\sum\limits_{m=1}^{\infty} \dfrac{{\Lambda}^m}{m}\Big( p_{m}\,[v_-]_{q^m}+{{ p}_{m}\,[v_+]_{q^m} + 2 \,[\beta]_{q^m} p_m\, {\widetilde  p_m}}\Big) \right) \ \ \ \ \ \ \ \ \ \ \ \ \ \ \ \ \ \ \ \ \ \ \ \ \ \ \ \ \ \ \ \ \ \ \ \ \ \ \ \ \ \ \ \
\end{align}
where in the last step we used the notations
\be
p_{m}=\sum\limits_{i=1}^{N_{+}} x_{i}^{m},\ \ \ {\widetilde  p_m}=\sum\limits_{j=1}^{N_{-}} y_{j}^{m},\ \ \  [v_{\pm}]_{q^m}=\dfrac{\ \ 1-q^{m v_{\pm}}}{1-q^m}=1+q^{m}+q^{2m}...+q^{(v_{\pm}-1)m}
\ee
Thus, one obtains
\begin{align}
I=\exp\left( \sum\limits_{m=1}^{\infty} \dfrac{[\beta]_{q^m}  {\Lambda}^m}{m} {\widetilde p}_{m}\Big( -{ p}_{m} - \dfrac{[v_+]_{q^m}}{[\beta]_{q^m}} \Big) \right)\,\exp\left( \sum\limits_{m=1}^{\infty} \dfrac{[\beta]_{q^m}  {\Lambda}^m}{m} p_{m}\Big( -{\widetilde p}_{m} - \dfrac{[v_-]_{q^m}}{[\beta]_{q^m}} \Big) \right)
\end{align}
Now to proceed to the expansion into a sum over the Young diagrams, we use the
Cauchy completeness formula for the MacDonald polynomials:
\be
\exp\Big( \sum\limits_{m=1}^{\infty} \dfrac{[\beta]_{q^m}}{m} p_{m} {\widetilde p}_{m} \Big)=\sum\limits_{R} \frac{C_{R}}{C_{R}^{\prime}}M_{R}(p_{m}) M_{R}( {\widetilde p}_{m})
\ee
Here $M(p_{m})$ are the normalized MacDonald polynomials, the hook lengths $C_{R}$ and
$C_{R}^{\prime}$ are defined by (\ref{hl}) and the summation goes over all Young
diagrams $R$. Using this, one finally obtains
\begin{align}
I=\sum_{A,B}\,{\Lambda}^{|A|+|B|}\,\frac{C_{A} C_{B}}{C_{A}^{\prime}C_{B}^{\prime}}\,M_{A}({\widetilde p}_m)\,M_{A}\Big(-p_m-   \dfrac{[v_+]_{q^m}}{[\beta]_{q^m}}   \Big)\,M_{B}(p_m)\,M_{B}\Big(-{\widetilde p}_m-   \dfrac{[v_-]_{q^m}}{[\beta]_{q^m}}   \Big)
\end{align}
Therefore, the $5d$ Dotsenko-Fateev integral takes the form:
\fr{\label{B5d}
B^{5D}(\Lambda)=\sum\limits_{A,B}\,{\Lambda}^{|A|+|B|}\,\frac{C_{A} C_{B}}{C_{A}^{\prime}C_{B}^{\prime}}\, \left<\ \ M_{A}\Big(-p_m-   \dfrac{[v_+]_{q^m}}{[\beta]_{q^m}}   \Big)\,M_{B}(p_m) \ \ \right>_{+} \,\left<\ \ M_{B}\Big(-{\widetilde p}_m-   \dfrac{[v_-]_{q^m}}{[\beta]_{q^m}}\Big) \,M_{A}({\widetilde p}_m)  \ \right>_{-}
}
This quantity has no the form of (\ref{tm}) and, therefore, does not factorize.
On the other hand, it avoids the problem of extra poles emerging in the Nekrasov
decomposition, see \cite{Dir}.

\subsection{The case of $\beta=1$}

The situation is completely different if $\beta=1$, when every double
average in (\ref{B5d}) factorizes and literally reproduces the corresponding Nekrasov
function which have no extra pole at $\beta=1$.
In this case, the MacDonald polynomials are reduced to the usual Schur
functions, however, the Selberg averages are still given by the
Jackson integrals $\left.M_{A}(p_{k})\right|_{\beta=1}=\chi_{A}(p_{k})$.
In order to calculate
\be
\label{dav}
\left.B^{5D}(\Lambda)\right|_{\beta=1}=\sum\limits_{A,B}\,{\Lambda}^{|A|+|B|} \left<\ \ \chi_{A}\Big(-p_m-   [v_+]_{q^m}   \Big)\,\chi_{B}(p_m) \ \ \right>_{+} \,\left<\ \ \chi_{B}\Big(-{\widetilde p}_m -[v_-]_{q^m}\Big) \,\chi_{A}({\widetilde p}_m)  \ \right>_{-}
\ee
one uses formula (\ref{tm}) of the Appendix which is reduced in this case to the form
\be
\label{aver}
\Big<\, \chi_{A}( p_k+[v]_k )\, \chi_{B}(p_k)\,\Big>=
\left(\prod\limits_{(i,j) \in A} q^{ i-1}\,\prod\limits_{(k,s) \in B} q^{ k-1+v}
\right)\dfrac{[v+N,A]_q\,[u+v+N,A]_q\,[u+N,B]_q\,[N,B]_q}{G^{+}_{A A}(0)
 G^{+}_{A^{\prime} B}(2N+u+v)  G^{-}_{B A^{\prime}}(2N+u+v) G^{+}_{B B}(0) }
\ee
where now
$$
[x,A]_q=\prod\limits_{(i,j)\in A} [x-i+j]_{q}, \ \ \ \textrm{and} \ \ \
G^{\pm}_{A B} (x) =\prod\limits_{(i,j)\in A}[x \pm \textrm{Arm}_{A}(i,j)
\pm \textrm{Leg}_{B}(i,j) \pm 1  ]_q
$$
Consider the double average appearing in (\ref{dav}):
\be
\nonumber
{\widetilde N}_{A,B}=\left<\ \ \chi_{A}\Big(-p_m- [v_+]_{q^m}   \Big)\,\chi_{B}(p_m) \ \ \right>_{+} \,\left<\ \ \chi_{B}\Big(-{\widetilde p}_m -[v_-]_{q^m}\Big) \,\chi_{A}({\widetilde p}_m)  \ \right>_{-}=\\ \nonumber
\\
=(-1)^{|A|+|B|}\,\left<\ \ \chi_{A^{\prime}}\Big( p_m +[v_+]_{q^m}   \Big)\,\chi_{B}(p_m)
 \ \ \right>_{+} \,\left<\ \ \chi_{B^{\prime}}\Big({\widetilde p}_m +[v_-]_{q^m}\Big)
\,\chi_{A}({\widetilde p}_m)  \ \right>_{-} \label{LFf}
\ee
where we used the formula for the characters of negative argument:
\be
\chi_{A}(-p)=(-1)^{|A|} \chi_{A^{\prime}}(p)
\ee
The usage of  (\ref{agt}) at the point $\epsilon_1+\epsilon_2$ gives
\be
v_+ + N_+ =-\mu_1-a,\ \ \ \ v_- + N_- =a-\mu_3\\ \nn
\\
u_++v_++N_+=-\mu_2-a,\ \ \ \ u_-+v_-+N_-=a-\mu_4\\ \nn
\\
u_+ + N_+ =\mu_1-a,\ \ \ \ u_- + N_- =a+\mu_3\\ \nn
\\
2 N_++u_++v_+=-2a,\ \ \ \ 2 N_-+u_-+v_-=2a\\ \nn
\\
N_+=\mu_2-a,\ \ \ \ \ N_-=\mu_4+a
\ee
Thus, (\ref{LFf}) takes the form
\be
{\widetilde N}_{A,B}=q^{(-\mu_1-\mu_2-2)|B|+(-\mu_3-\mu_4-2)|A|} \prod\limits_{(i,j)\in A} q^{i+j}
\prod\limits_{(i,j)\in B} q^{i+j} \times \nn \\  \times
\dfrac{\prod\limits_{k=1}^{2} [-\mu_k-a,A^{\prime}]_{q} [\mu_k-a,B]_{q}
\prod\limits_{k=3}^{4} [\mu_k+a,A]_{q} [-\mu_k+a,B^{\prime}]_{q}
  }{G_{A^{\prime} A^{\prime}}^{+}(0)G_{A B}^{+}(-2a)G_{B A}^{-}(-2a)
G_{B B}^{+}(0)
G_{B^{\prime} B^{\prime}}^{+}(0)G_{B A}^{+}(2a)G_{A B}^{-}(2a)
G_{A A}^{+}(0)
}
\ee
Note that at $\beta=1$ all the factors here can be expressed through the functions
(\ref{f}) and (\ref{g}):
\be
[x,A]=f_{A}^{+}(x),\ \ \ [-x,A^{\prime}]=f_{A}^{-}(x),\ \ \ g_{A,B}(x)=G_{A,B}^{+}(x) G_{A,B}^{-}(-x)
\ee
and one can reduce the expression to the Nekrasov functions (\ref{Shf}).
 Finally, with the use of the
following simple identities:
\be
\prod\limits_{(i,j)\in A} q^{\textrm{Leg}_{i,j}(A)}=\prod\limits_{(i,j)\in A} q^{j-1}\\
\prod\limits_{(i,j)\in A} q^{\textrm{Arm}_{i,j}(A)}=\prod\limits_{(i,j)\in A} q^{i-1}\\
\prod\limits_{(i,j)\in A} q^{B_{i}}=\prod\limits_{(i,j)\in B} q^{A_i}
\ee
one finds
\be
\label{LF}
\begin{array}{|c|}
\hline \\
{\widetilde N}_{A,B}= N_{A,B} \\
\\
\hline
\end{array}
\ee
where $N_{A,B}$ is the Nekrasov function defined by (\ref{Shf}) and restricted to
 $\epsilon=\epsilon_1+\epsilon_2 =0$.
Therefore, finally we arrive at
\be
\left. B^{5D}(\Lambda)\right|_{\beta=1} =\sum\limits_{A,B}\, \left.N_{A,B}\right|_{\epsilon_1+
\epsilon_2=0} {\Lambda}^{|A|+|B|} = \left.Z_{Nek}^{5D}(\Lambda)\right|_{\epsilon_1+\epsilon_2=0}
\ee

\section*{Acknowledgements}

Our work is partly supported by Ministry of Education and Science of
the Russian Federation under contract 14.740.11.081 (A.Mir., A.Mor., Sh.Sh.)
and 14.740.11.0347 (A.S.), by RFBR
grants 10-02-00509 (A.Mir.), 10-02-00499 (A.Mor.\& Sh.Sh.) and 09-02-00393 (A.S.),
by joint grants 11-02-90453-Ukr, 09-02-93105-CNRSL, 09-02-91005-ANF,
10-02-92109-Yaf-a, 11-01-92612-Royal Society.

\section*{Appendix}

\subsection*{MacDonald polynomials}

\bigskip
\noindent \textbf{Definition.}
The MacDonald polynomials is the distinguished basis in the space
 of symmetric polynomials of $\{x_i\}$. Let us first define the basis
 \be
p_R=p_{R_1}(x)\ldots p_{R_n}(x)=p_1^{m_1}(x)p_2^{m_2}(x)\ldots
 \ee
 where
\be p_{k}=\sum\limits_{i=1}^{N} x_{i}^{k} \ee
with the scalar product
\be
\langle p_R|p_{R'}\rangle=\delta_{RR'}\prod_km_k!k^{m_k}
\prod_{i=1}^n{1-q^{R_i}\over 1-t^{R_i}},\ \ \ \ \ \ \ t=q^{\beta}
\ee
which can be also manifestly realized by
\be\label{orth}
\left.\langle f(p_k)|g(p_k)\rangle=f\left(k{1-q^k\over 1-t^k}
{\partial\over\partial p_k}\right)
g(p_k)\right|_{p_k=0}
\ee
Introduce the symmetric functions
$m_R=\sum_\sigma x_1^{R_{\sigma (1)}}x_2^{R_{\sigma (2)}}...$
with $R_i$ being the lengths of rows of the Young diagram $R$ and
the (partial) ordering of the Young diagrams is defined as $R\ge R'$ iff
$|R|=|R'|$ and $\sum_{k=1}^iR_k\ge\sum_{k=1}^iR'_k$ for all $i$. Then,
the MacDonald polynomials are the polynomials given by the expansion\footnote{We omit
the superscript $q,t$ unless this may lead to a confusion.}
\be
M_R^{q,t}(x_1,...,x_n)=\sum_{R'<R}c_{RR'}m_{R'}=m_R+\ldots
\ee
with the unit coefficient $c_{RR}$ that satisfy the orthogonality condition
\be
\langle M_R^{q,t}|M_{R'}^{q,t}\rangle=0\ \ \ \ \ \ \ \hbox{if } R\ne R'
\ee
\bigskip
\noindent \textbf{Examples.}
The few first MacDonald polynomials are:
\begin{small}
$$
M_{1}=p_{{1}},\ \ \ \ \ M_{2}=\frac{(1-t)(1+q)}{(1-tq)}\frac{p_1^2}{2}+\frac{(1+t)(1-q)}{(1-tq)}\frac{p_2}{2}, \ \ \ \ \ M_{11}=\frac{p_1^2}{2}-\frac{p_2}{2}\ \ \ \ \ \ \ \ \ \ \ \ \ \  \ \ \  \ \ \ \ \  \ \ \ \
 \ \ \ \ \ \  \ \ \
$$
$$
M_{3}=\frac{(1+q)(1-q^3)(1-t)^2}{(1-q)(1-tq)(1-tq^2)}\frac{p_1^3}{6}+\frac{(1-t^2)(1-q^3)}{(1-tq)(1-tq^2)}\frac{p_1 p_2}{2}+\frac{(1-q)(1-q^2)(1-t^3)}{(1-t)(1-t q) (1-t q^2)}\frac{p_{3}}{3} \ \ \ \ \ \  \ \ \ \ \ \ \ \  \ \ \ \ \ \ \ \
$$
$$
M_{21}=\frac{(1-t)(2qt+q+t+2)}{1-qt^2} \frac{p_{1}^3}{6}+\frac{(1+t)(t-q)}{1-qt^2}\frac{p_1p_2}{2} -\frac{(1-q)(1-t^3)}{(1-t)(1-qt^2)}\frac{p_{3}}{3},\ \ \ M_{111}=\frac{p_{1}^3}{6} -\frac{p_{1}p_{2}}{2}+\frac{p_{3}}{3}
$$
\end{small}
\bigskip
\noindent \textbf{Limiting cases.}
At the point $t=q$ ($\beta=1$) the MacDonald polynomials reduces to the Schur polynomials:
\be
\left.M(x_{i})\right|_{t=q}=\chi_R(x_{i})=
\frac{\det_{1\leq i,j\leq N} x_i^{R_j+N-j}}{\Delta(x)}=
\det_{ij}
S_{R_i - i+j}(p)
\ee
where $\exp\left(\sum
p_kz^k/k\right) = \sum_k S_k(t)z^k$
and the Van-der-Monde determinant $\Delta(x)  = \det_{ij} x_i^{N-j} =
 \prod_{i<j}^N(x_i-x_j)$.

In the intermediate case $q=1$ the MacDonald polynomials degenerate to the symmetric Jack
polynomials which are relevant for the proof of AGT conjecture in $4d$ case :
\be
\left.M(x_{i})\right|_{q=1}=J^{\beta}(x_{i})
\ee
\bigskip
\noindent \textbf{MacDonald polynomials as a set of eigenfunctions.}
They are also uniquely defined as the common system of eigenfunctions of the commuting
set of operators, which are nothing but the Ruijsenaars Hamiltonians \cite{Rui,5dB}:
\be\label{Ham}
\hat H_{k}=\sum\limits_{i_{1}<...<i_{k}} {1\over\Delta (x)} \hat T_{i_{1}}...\hat T_{i_{k}}
\Delta (x)\, \hat Q_{i_{1}}...\hat Q_{i_{k}},\ \ \ \ [\hat H_{k},\hat H_{m}]=0
\ee
where the shift operators are defined as:
\be
\hat T_{k}=q^{\beta x_{k} \partial_{x_{k}}}, \ \ \ \hat Q_{k}=q^{(1-\beta) x_{k}\partial_{x_{k}}}
\ee
The spectrum of (\ref{Ham}) can be defined from the eigenvalues of spectral operator:
\be
\left(\sum\limits_{k=0}^{n} z^{k} \hat H_{k} \right)M_{R}(x_{1},...,x_{n})=
\prod\limits_{i=1}^{\infty} (1+z\, q^{R_{i}+\beta(n-i)}) M_{R}(x_{1},...,x_{n})
\ee
Note that at $\beta=1$, when $\hat Q_{k}=1$ the spectral operator can be summed exactly:
\be
\sum\limits_{k=0}^{n} z^{k} \hat H_{k}|_{t=q}=
\sum\limits_{k=0}^{n} z^{k}\sum\limits_{i_{1}<...<i_{k}} {1\over\Delta (x)}
\hat T_{i_{1}}...\hat T_{i_{k}} \Delta (x)={1\over\Delta (x)} \prod\limits_{k=1}^{n}
(1+z \hat T_{k})\, \Delta (x)
\ee
and one obtains
\be
\left[{1\over\Delta (x)} \prod\limits_{k=1}^{n}(1+z \hat T_{k})\, \Delta (x)\right] \chi_{R}(x)
=\prod\limits_{i=1}^{n} (1+z q^{n-i+R_{i}})\chi_{R}(x)
\ee
\bigskip
\noindent \textbf{Orthogonality.}
Besides the scalar product (\ref{orth}), there is another scalar product
$<,>^*$ such that the
MacDonald polynomials are also orthogonal w.r.t. it, but have other norms.
This scalar product is given by the integral with the MacDonald measure:
\be
\left< f, g \right>^*=\oint\limits_{|z_1=1|}\frac{dz_1}{z_1}...\oint\limits_{|z_{N}|=1}\frac{dz_N}{z_N} \prod\limits_{m=0}^{\beta-1} \prod\limits_{i\neq j} \Big( 1-q^m \frac{z_{i}}{z_{j}} \Big) f(z_{1},...,z_{N}) g(z_{1}^{-1},...,z_{N}^{-1})
\ee
and the normalization condition is
\be
\left< M_{A}, M_{B} \right>^*=\delta_{A,B}\, \frac{C_{A}^{\prime}}{C_{A}} \frac{[\beta N, A]}{[\beta N+1-\beta, A]}
\ee
with the $q$-Pochhammer symbol
\be\label{qPoch}
 [x,A]_q=\prod\limits_{(i,j)\in A} [x-i\beta+j+\beta-1]_q,
\ee
and
\be \label{hl} C_{A}^{\prime}=\prod\limits_{(i,j)\in A}\,[\beta \textrm{Arm}_{A}(i,j)+
\textrm{Leg}_{A}(i,j) +\beta ]_q,\ \ \ C_{A}=\prod\limits_{(i,j)\in A}\,[\beta \textrm{Arm}_{A}(i,j)+\textrm{Leg}_{A}(i,j) +1 ]_q\ee
\bigskip
\noindent \textbf{Cauchy-Stanley completeness identity.}
The MacDonald polynomials satisfy the following identity of expansion of the bilinear exponential:
\be
\label{ls}
\begin{array}{|c|}
\hline\\
\exp\Big( \sum\limits_{k=1}^{\infty} \dfrac{[\beta]_{q^k}}{k} p_{k} {\widetilde{p}}_{k}\Big)
=\sum\limits_{R} \frac{C_{R}}{C_{R}^{\prime}} M_{R}(p_{k}) M_{R}({\widetilde{p}}_{k})\\
\\
\hline
\end{array}
\ee
A few different representations of this identity are known in the literature, all of them can
be obtained from (\ref{ls}) by simple algebraic manipulations. For example, with
$p_{k}=\sum_{i} x_{i}^k$, ${\widetilde{p}}_{k}=\sum_{j} y^{k}_{j}$
the l.h.s. of (\ref{ls}) can be rewritten as follows:
\be
\exp\Big( \sum\limits_{k=1}^{\infty} \dfrac{[\beta]_{q^k}}{k} p_{k} {\widetilde{p}}_{k}\Big)
=\exp\Big( \sum\limits_{i,j}\sum\limits_{k=1}^{\infty} \dfrac{1-t^k}{k (1-q^k)}  x_{i}^{k} y_{j}^{k} \Big)=\prod\limits_{i,j} \dfrac{\exp\Big( -\textrm{Li}_2(t x_{i} y_{j}|q) \Big)}{\exp\Big(-\textrm{Li}_2(x_{i} y_{j}|q)\Big)}
\ee
where $\textrm{Li}_{2}(x|q)$ is the quantum dilogarithm function:
\be
\textrm{Li}_{2}(x|q)=\sum\limits_{k=1}^{\infty} \dfrac{x^{k}}{k (1-q^k)}
\ee
Using the identity for the quantum dilogarithm, which relates it with the $q$-exponential
\be
\exp\Big(-\textrm{Li}_{2}(x|q)\Big)=\prod\limits_{k=0}^{\infty} (1-q^k x)
\stackrel{\textrm{def}}{=}(x;q)_{\infty}=
\sum\limits_{n=0}^{\infty} \dfrac{(-1)^n  }{[n]_q!(1-q)^n} q^{n(n-1)/2} x^n \stackrel{\textrm{def}}{=} \textrm{E}_{q}(-x)
\ee
one obtains the Cauchy completeness identity in the infinite product form
or, equivalently, in the  $q$-exponential form:
\be
\label{ls2}
\begin{array}{|c|}
\hline\\
\sum\limits_{R} \frac{C_{R}}{C_{R}^{\prime}} M_{R}(p_{k}) M_{R}({\widetilde{p}}_{k})=
\prod\limits_{i,j}\dfrac{(t x_{i}y_{j})_{\infty}}{(x_{i}y_{j})_{\infty}}=
\prod\limits_{i,j} \dfrac{\textrm{E}_{q}(-t x_{i} y_{j})}{\textrm{E}_{q}(- x_{i} y_{j})}\\
\\
\hline
\end{array}
\ee

Finally, consider (\ref{ls}) at the point ${\widetilde{p}}_{k}=-{\widetilde{p}}_{k}/[\beta]_{q^k}$:
\be
\label{e1}
\exp\Big( -\sum\limits_{k=1}^{\infty} \frac{p_{k}{\widetilde{p}}_{k}}{k}  \Big)=\sum\limits_{R} \frac{C_{R}}{C_{R}^{\prime}}M_{R}(p_{k}) M_{R}(-{\widetilde{p}}_{k}/[\beta]_{q^k})
\ee
Expressing the l.h.s. of this identity through the eigenvalues $(\ref{ev})$, one obtains
\be
\label{e2}
\exp\Big( -\sum\limits_{k=1}^{\infty} \frac{p_{k}{\widetilde{p}}_{k}}{k}  \Big)=\prod\limits_{i,j} \exp\Big( -\sum\limits_{k=1}^{\infty} \frac{x_{i}^k y_{j}^k}{k}  \Big)=\prod\limits_{i,j} \exp\Big( \ln( 1-x_{i} y_{j})  \Big)=\prod\limits_{i,j} (1-x_{i} y_{j})
\ee
The r.h.s. can be transformed by utilizing the identity for the MacDonald polynomial of negative
argument
\be\label{inv}
M_{R}^{q,t}\Big(-\frac{p_{k}}{[\beta]_{q^k}}\Big)=(-1)^{|R|} \frac{C_{R}^{\prime}}{C_{R}} M^{t,q}_{R^{\prime}}(p_{k})
\ee
where $R^{\prime}$ stands for the transposed Young diagram (conjugated representation)
and we write the deformation parameters $q$ and $t$ explicitly to emphasize that the
MacDonald polynomials at the r.h.s. and l.h.s. of this identity are calculated at
interchanged $t$ and $q$. One can easily check that (\ref{inv}) provides an involution
transformation by applying it twice which results into unity. In order to proof, one
suffices to note that
\be
C'_A(\beta)=\beta^{|A|}C_{A'}\left({1\over\beta}\right),\ \ \ \ \ [\beta]_{q^k}[\beta^{-1}]_{t^k}=1
\ee
Applying this involution transformation to (\ref{e1}) and using (\ref{e2}) one gets
\be
\sum\limits_{R} (-1)^{|R|} M_{R}^{q,t}(p_{k}) M_{R^{\prime}}^{t,q}(\tilde p_{k})=\prod\limits_{i,j}(1-x_{i} y_{j})
\ee
Switching again to the eigenvalues and using that $M_{R}(-y_{j})=(-1)^{|R|} M_{R}(y_{j})$
one finally obtains the standard form of the Cauchy completeness identity:

\be
\label{ls4}
\begin{array}{|c|}
\hline\\
\sum\limits_{R} M_{R}^{q,t}(x_{i}) M_{R^{\prime}}^{t,q}(y_{j})=\prod\limits_{i,j}(1+x_{i} y_{j})\\
\\
\hline
\end{array}
\ee
\bigskip

\subsection*{$q$-deformed $\beta$-ensembles}

We consider the following average for the polynomial $f(x_{1},...,x_{N})$:
\begin{align}
\Big< f \Big> \ = \ \dfrac{1}{S}
\int\limits_{0}^{1} d_qx_1 \ldots \int\limits_{0}^{1} d_qx_{N} \prod\limits_{i\neq j} \prod\limits_{k=0}^{\beta-1} (x_i - q^k\,x_j) \prod\limits_{i} x_i^{u} \prod\limits_{k=0}^{v-1}(1-q^k \,x_i) \ f\big(x_1, \ldots, x_{N}\big)
\label{xAverage}
\end{align}
where the normalization
\begin{align}
S= \int\limits_{0}^{1} d_qx_1 \ldots d_qx_{N} \prod\limits_{i\neq j} \prod\limits_{k=0}^{\beta-1} (x_i - q^k\,x_j) \prod\limits_{i} x_i^{u}\prod\limits_{k=0}^{v-1} (1-q^k\,x_i)
\end{align}
provides $\Big< 1 \Big> = 1$. Here we use the notion of Jackson integral:
\begin{align}
\label{ji}
\int_0^a f(x) d_{q} x=(1-q) a \sum\limits_{k=0}^{\infty} q^k f( q^k a ), \ \ \ \textrm{in particular} \ \ \ \int\limits_{0}^{1} f(x) d_{q} x=(1-q) \sum\limits_{k=0}^{\infty} q^k f( q^k )
\end{align}
The Jackson integrals of polynomials are equal to
$$
\int\limits_{0}^{1} x^n d_{q} x=\dfrac{1}{[n+1]_q},\ \ \ [n]_{q}=\dfrac{1-q^n}{1-q}=1+q+...+q^{n-1} $$
The average (\ref{xAverage}) is the obvious $q$-deformation of the Selberg $\beta$-ensemble
considered in our previous paper \cite{Dir}:
\begin{align}
\Big< f \Big>^{{\rm Selb}} \ = \ \dfrac{
\int\limits_{0}^{1} dx_1 \ldots \int\limits_{0}^{1} dx_{N} \prod\limits_{i<j} (x_i - x_j)^{2 \beta} \prod\limits_{i} x_i^{u} (x_i - 1)^{v} \ f\big(x_1, \ldots, x_{N}\big)}{\int\limits_{0}^{1} dx_1 \ldots \int\limits_{0}^{1} dx_{N} \prod\limits_{i<j} (x_i - x_j)^{2} \prod\limits_{i} x_i^{u} (x_i - 1)^{v} }
\end{align}
For the sake of simplicity, we keep in (\ref{xAverage}) the parameters $\beta$ and $v$ integer,
extension to non-integer values of the parameters being straightforward.
For instance, the MacDonald measure in (\ref{xAverage})\footnote{Note that
(\ref{Ham}) involves the ordinary Van-der-Monde determinant, not (\ref{MCV}).}
\be\label{MCV}
\Delta^{MC}(x_{i})=\prod\limits_{i\neq j} \prod\limits_{m=0}^{\beta-1} (x_{i}- q^{m} x_{j})
\ee
can be rewritten in the form:
\be
\label{nib}
\Delta^{MC}(x_{i})=\prod\limits_{i\neq j} \prod\limits_{m=0}^{\infty} \left( \dfrac{x_{i}-q^m x_{j}}{x_{i}-t q^m x_{j}} \right)=\prod\limits_{i\neq j} \exp\left(-\sum\limits_{k=1}^{\infty} \dfrac{1}{k} \dfrac{1-t^k}{1-q^k} \Big(\frac{x_{j}}{x_{i}}\Big)^k\right),\ \ \ \ t=q^{\beta}
\ee
where $\beta$ can take non-integer values. Analogously, at non-integer $v$
\be
\label{niv}
\prod\limits_{k=0}^{v-1} (1-q^k x)\longrightarrow\exp\left( -\sum\limits_{m=1}^{\infty} \dfrac{1}{m} \dfrac{1-q^{v m}}{1-q^m} x^m \right)
\ee

\subsection*{1-MacDonald average \label{ap}}

The average of the single MacDonald polynomial in the $q$-deformed $\beta$-ensemble,
which generalizes the celebrated Kadell formula \cite{Kadell}, has the form
\fr{
\label{om}
\Big< M_A(p) \Big>=q^{W_A(v,\beta)}\,\dfrac{[N \beta, A]_q  [u+N \beta+1-\beta,A]_q}{d_q(A) [u+v+2 N \beta +2-2\beta,A]_q}
}
where
\begin{align}
q^{W_Y (v,\beta)}=\prod\limits_{(i,j)\in Y} q^{v+(i-1) \beta }= q^{|Y| v} \prod\limits_{i=1}^{h(A)} q^{(i-1) \beta Y_{i}}
\end{align}
and
\begin{align}
\label{dA}
d_{q}(Y)=\prod\limits_{(i,j)\in Y}\,[\beta+(Y_i-j)+\beta(Y_{j}^{\prime}-i)]_q
\end{align}
In the case of Jack polynomials this latter quantity could be presented as a particular
value of the polynomial:
\be
J_{A}\Big(p_{k}=\delta_{k,1}\Big)=\dfrac{\beta^{|A|}}{\prod\limits_{(i,j)\in Y}\Big(\beta+(Y_i-j)+\beta(Y_{j}^{\prime}-i)\Big)}
\ee
which led to formula (74) in \cite{Dir}
(there was a misprint in \cite{Dir}):
\begin{align}
\Big< J_{A}(p_{k}) \Big>^{{\rm Selb}}= J_A(\delta_{k,1} ) \dfrac{[N \beta, A]  [u+N \beta+1-\beta,A] }{ \beta^{|A|}  [u+v+2 N \beta +2-2\beta,A]}
\end{align}
However, in the $q$-deformed case there is no such a simple relation:
$$
M_A(\delta_{k,1}) \neq \dfrac{\beta^{|A|}}{d_q(A)}
$$

\subsection*{2-MacDonald average}

We have found the following formula for the Selberg average of product of two non-normalized
MacDonald polynomials:
\fr{\nonumber
\Big< M_{A}(p_{k}+w_{k}) M_{B}(p_{k}) \Big>=
q^{W_{A,B}(v,\beta)}
\,\dfrac{[v+N\beta+1-\beta,A]_{q}[u+N\beta+1-\beta,B]_q}{[N\beta,A]_q[u+v+N\beta+2-2\beta,B]_q}\times
\\
\\
\times \dfrac{\prod\limits_{i,j=1}^{N}\,P_{\beta}\Big( u+v+2\beta N +2-\beta(1+i+j) \Big)}{\Big(\prod\limits_{1\leq i<j \leq N}\,P_{\beta}(\beta j-\beta i)\ \ \Big)^{2}}
\dfrac{ \prod\limits_{1\leq i<j \leq N}\,P_{\beta}\Big( A_{i}-A_{j}+\beta(j-i) \Big)\,\prod\limits_{1\leq i<j \leq N}\,P_{\beta}\Big( B_{i}-B_{j}+\beta(j-i) \Big) }{ \prod\limits_{i,j=1}^{N}\,P_{\beta}\Big( u+v+2\beta N +2+A_{i}+B_{j}-\beta(1+i+j) \Big) }
\label{tm}
}
Note that this expression explicitly depends on $N$, the number of parameters $x_i$ in
(\ref{xAverage}), and we use the rule $A_i=0$ if $i$ exceeds the number of rows in $A$.
Note that in our normalization $\langle 1\rangle =1$ for the
empty Young diagrams $M_{[]}(p_k)=1$:
\be
\Big< M_{[]}(p_{k}+w_{k}) M_{[]}(p_{k}) \Big>=\langle 1\rangle =1
\ee
In formula (\ref{tm})
\be\label{pb}
P_{\beta}( x ) =\dfrac{\Gamma_{q}(x+\beta)}{\Gamma_{q}(x)}=
\prod\limits_{k=0}^{\beta-1} [x+k]_q
\ee
the latter identity being correct in the case of integer $\beta$. At last,
\be
\label{shift}
w_k=-q^{v k} \dfrac{[\beta-v-1]_{q^{k}}}{[\beta]_{q^k}}= -q^{v k} \dfrac{[(\beta-v-1)k]_{q}}{[\beta k]_{q}}
\ee
since
$$
[n]_{q^k}=\dfrac{[nk]_q}{[k]_q}
$$
Note that the main feature of (\ref{tm}), its complete factorization into $q$-number
factors, happens only at these specific values of $w_k$.

\subsection*{Example}

We now illustrate the use of these formulas in
the simplest example of the average
$<p_1+w_1>$. It can be considered as
$<M_1(p)>+w_1$ and evaluated with the help of
(\ref{om}), or as $<M_1(p+w)M_0(p)>$ and
evaluated with the help of (\ref{tm}).

In the first case one has:
\be
<p_1+w_1> \ \stackrel{(\ref{om})}{=}\
w_1 + q^v\frac{[N\beta]_q}{[\beta]_q}
\frac{[u+N\beta+1-\beta]_q}{[u+v+2N\beta+2-2\beta]_q}
= \nn \\ \\
=\left\{\begin{array}{ccc}
w_1+q^{-\mu_1-\mu_2}\frac{[\mu_2-a]_q[\mu_1-a]_q}
{[\beta]_q[-2a+1-\beta]_q} & & {\rm for}\ \ <\ldots>_+\ \
\textrm{in} \ (\ref{agt}) \\ && \\
w_1+q^{-\mu_3-\mu_4}\frac{[\mu_4+a]_q[\mu_3+a]_q}
{[\beta]_q[-2a+1-\beta]_q} & & {\rm for}\ \ <\ldots>_-\ \
\textrm{in}\ (\ref{agt})
\end{array}\right.
\ee
These expressions are nicely decomposed into a product
of two "linear" factors for $w_1=0$ and also for
\be
w_1 = \left\{\begin{array}{c}
-q^{-\mu_1-\mu_2}\frac{[\beta-1+\mu_1+\mu_2]_q}{[\beta]_q}\\ \\
-q^{-\mu_3-\mu_4}\frac{[\beta-1+\mu_3+\mu_4]_q}{[\beta]_q}
\end{array}\right.
\ee
This distinguished value of $w_1$ is especially easy
to find for $q=1$: the discriminant of quadratic polynomial
$(a-\mu_1)(a-\mu_2) + \beta w_1(-2a+1-\beta)$
is the full square:
\be
\nn
D = (\mu_1+\mu_2+2\beta w_1)^2 - 4\Big(\mu_1\mu_2+\beta w_1(1-\beta)\Big)
= (\mu_1-\mu_2)^2 + 4\beta w_1\Big(\beta w_1 + \mu_1+\mu_2-(1-\beta)\Big)
= \\= (\mu_1-\mu_2)^2 \ \ \ \ \ \
{\rm for} \ \ \ \ w_1=0 \ \ \ {\rm or}\ \ \
w_1 = \frac{-\mu_1-\mu_2+1-\beta}{\beta} \ \ \ \ \ \ \ \ \ \ \ \ \ \ \ \ \ \ \
\ee
When $q$ is switched on, one has:
\be
\frac{1}{[\beta]_q}\!\!\left(q^{-\mu_1-\mu_2}\frac{[\mu_1-a]_q
[\mu_2-a]_q}{[-2a+1-\beta]_q} - q^{-\mu_1-\mu_2}
[\beta-1+\mu_1+\mu_2]_q\right)
= q^{\beta-1}\frac{[1-\beta-\mu_1-a]_q[1-\beta-\mu_2-a]_q}
{[\beta]_q[1-\beta-2a]_q}
\label{p+w_av}
\ee
Note that the main role of the $w$-shift is to change the relative sign
between $a$ and $\mu$ in the numerator, like in (\ref{twodeco}).
However, the value of this shift, which is important for
factorization property, is here different from the value of the shift
in (\ref{B5d}), needed to reproduce the conformal block:
the shifts are the same only for $\beta=1$.

In the second representation of the same average one uses
formula (\ref{tm}) with $A=[1]$ and $B=[]$.
In this case the products of $P_{\beta}$-factors
get non-trivial contributions only from $i=1$:
\be
\Big<p_1+w_1 \Big>=q^{\beta-1}\dfrac{[v+N\beta+1-\beta]_q}{[N\beta]_q}\prod\limits_{j=2}^{N}
\dfrac{P_{\beta}\big(1+\beta(j-1)\big)}{P_\beta\big(\beta(j-1)\big)}\prod\limits_{j=1}^{N}
\dfrac{P_{\beta}(u+v+2 \beta N+2-\beta(2+j))}{P_{\beta}(u+v+2\beta N+3-\beta(2+j))}\,
\ee
Using the property of $P_{\beta}(x)$:
$$
\dfrac{P_{\beta}(x+1)}{P_{\beta}(x)}=\dfrac{[x+\beta]_q}{[x]_q}
$$
which is obvious from its definition (\ref{pb}), we find
\be
\nn
\Big<p_1+w_1\Big>=q^{\beta-1} \dfrac{[v+N\beta+1-\beta]_q}{[N\beta]_q}
\prod\limits_{j=2}^{N} \dfrac{[\beta j]_q}{[\beta (j-1)]_q}
\prod\limits_{j=1}^{N} \dfrac{[u+v+2 \beta N+2-\beta (j -2) ]_q}{[u+v+2\beta N+2-\beta (j -1)]_q}\,
=
\ee
\be
\nn
=q^{\beta-1} \dfrac{[N\beta]_q}{[\beta]_q}\,\dfrac{[u+v+\beta N +2  -2\beta]_q}{[u+v+2\beta N +2-2\beta]_q}\,
\dfrac{[v+N\beta+1-\beta]_q}{[N\beta]_q}=
q^{\beta-1} \,\dfrac{[u+v+\beta N +2  -2\beta]_q [v+N\beta+1-\beta]_q }{[\beta]_q
[u+v+2\beta N +2-2\beta]_q}=
\ee
\be
= q^{\beta-1} \,\dfrac{[1-\beta-\mu_2-a]_q [1-\beta-\mu_1-a]_q }{[\beta ]_q [1-\beta-2 a ]_q}
\ee
where at the last stage we substituted parameters (\ref{agt}) for the $<\ldots>_+$ average.
The result is exactly the same as (\ref{p+w_av}).


\begin{thebibliography}{12}

\bibitem{AGT} L.Alday, D.Gaiotto and Y.Tachikawa,
Lett.Math.Phys. {\bf 91} (2010) 167-197, arXiv:0906.3219

\bibitem{AGTmore}
N.Wyllard,
JHEP {\bf 0911} (2009) 002, arXiv:0907.2189;
arXiv:1011.0289;
arXiv:1012.1355;\\
N.Drukker, D.Morrison and T.Okuda, JHEP {\bf 0909} (2009) 031, arXiv:0907.2593;\\
S.Iguri and C.Nunez, JHEP {\bf 11} (2009) 090 , arXiv:0908.3460;\\
D.Nanopoulos and D.Xie, arXiv:0908.4409;
JHEP {\bf 1003} (2010) 043, arXiv:0911.1990;
arXiv:1005.1350;
arXiv:1006.3486;\\
L.Alday, D.Gaiotto, S.Gukov, Y.Tachikawa and H.Verlinde,
JHEP {\bf 1001} (2010) 113, arXiv:0909.0945;\\
N.Drukker, J.Gomis, T.Okuda and J.Teschner,
JHEP {\bf 1002} (2010) 057, arXiv:0909.1105;\\
A.Marshakov, A.Mironov and A.Morozov,
JHEP 11 (2009) 048, arXiv:0909.3338;\\
R.Poghossian,
JHEP {\bf 0912} (2009) 038,  arXiv:0909.3412;\\
A.Gadde, E.Pomoni, L.Rastelli and S.Razamat, JHEP {\bf 1003} (2010) 032, arXiv:0910.2225;\\
L.Alday, F.Benini and Y.Tachikawa, Phys.Rev.Lett. {\bf 105} (2010) 141601, arXiv:0909.4776;\\
S.Kanno, Y.Matsuo, S.Shiba and Y.Tachikawa,
Phys.Rev. {\bf D81} (2010) 046004, arXiv:0911.4787;\\
G.Bonelli and A.Tanzini,
arXiv:0909.4031;\\
J.-F.Wu and Y.Zhou,
arXiv:0911.1922;\\
G.Giribet,
 JHEP {\bf 01} (2010) 097, arXiv:0912.1930;\\
V.Alba and And.Morozov,
Nucl.Phys. {\bf B840} (2010) 441-468, arXiv:0912.2535;\\
M.Fujita, Y.Hatsuda, Y.Koyama and T.-Sh.Tai,
JHEP {\bf 1003} (2010) 046, arXiv:0912.2988;\\
M.Taki,
arXiv:0912.4789;
arXiv:1007.2524;\\
Piotr Sulkowski,
JHEP {\bf 1004} (2010) 063, arXiv:0912.5476; arXiv:1012.3228\\
N.Nekrasov and E.Witten, arXiv:1002.0888;\\
R.Santachiara and A.Tanzini,
arXiv:1002.5017;\\
S.Yanagida, arXiv:1003.1049;
arXiv:1010.0528;\\
N.Drukker, D.Gaiotto and J.Gomis
arXiv:1003.1112;\\
F.Passerini,
JHEP {\bf 1003} (2010) 125, arXiv:1003.1151;\\
C.Kozcaz, S.Pasquetti and N.Wyllard,
arXiv:1004.2025;\\
S.Kanno, Y.Matsuo and S.Shiba,
arXiv:1007.0601;\\
H.Awata, H.Fuji, H.Kanno, M.Manabe and Y.Yamada,
arXiv:1008.0574;\\
C.Kozcaz, S.Pasquetti, F.Passerini and N.Wyllard,
arXiv:1008.1412;\\
H.Itoyama, T.Oota and N.Yonezawa,
arXiv:1008.1861;\\
A.Braverman, B.Feigin, M.Finkelberg and L.Rybnikov, arXiv:1008.3655;\\
Ta-Sheng Tai,
arXiv:1006.0471;
arXiv:1008.4332;
arXiv:1012.4972;\\
M.Billo, L.Gallot, A.Lerda and I.Pesando,
arXiv:1008.5240;\\
A.Brini, M.Marino and S.Stevan,
arXiv:1010.1210;\\
M.C.N.Cheng, R.Dijkgraaf adn C.Vafa,
arXiv:1010.4573;\\
Y.Yamada,
arXiv:1011.0292;\\
J.-F. Wu,
arXiv:1012.2147;\\
A.Marshakov, arXiv:1101.0676;\\
G.Bonelli, A.Tanzini and J.Zhao, arXiv:1102.0184;\\
A.Belavin and V.Belavin, arXiv:1102.0343;\\
A.Gorsky, arXiv:1102.1841;\\
O.P.Santillan, arXiv:1103.1422;\\
H.Itoyama and N.Yonezawa, arXiv:1104.2738;\\
G.Bonelli, K.Maruyoshi and A.Tanzini, arXiv:1104.4016;\\
H.Kanno and Y.Tachikawa, arXiv:1105.0357;\\
M.Aganagic, M.C.N.Cheng, R.Dijkgraaf, D.Krefl and C.Vafa,  arXiv:1105.0630

\bibitem{AGTmmm}
A.Marshakov, A.Mironov and A.Morozov,  Theor.Math.Phys. {\bf 164} (2010) 831-852
(Teor.Mat.Fiz.164:3-27,2010), arXiv:0907.3946

\bibitem{power}
A.Mironov and A.Morozov, Phys.Lett. {\bf B680} (2009) 188-194, arXiv:0908.2190

\bibitem{2pap}
A.Mironov and A.Morozov, Nucl.Phys. {\bf B825} (2009) 1-37, arXiv:0908.2569;
Phys.Lett. {\bf B682} (2009) 118-124, arXiv:0909.3531

\bibitem{MMMM}
Andrey Mironov, Sergey Mironov, Alexei Morozov
and Andrey Morozov,
Theor.Math.Phys. 165 (2010) 1662-1698 (Teor.Mat.Fiz. 165 (2010) 503-542), arXiv:0908.2064

\bibitem{nonconf}
D.Gaiotto, arXiv:0908.0307;\\
A.Marshakov, A.Mironov and A.Morozov,
Phys.Lett. {\bf B682} (2009) 125-129, arXiv:0909.2052;\\
V.Alba and And.Morozov,
JETP Lett. {\bf 90} (2009) 708-712 , arXiv:0911.0363

\bibitem{NS} N.Nekrasov and S.Shatashvili, arXiv:0908.4052;\\
A.Mironov and A.Morozov, 
JHEP {\bf 04} (2010) 040, arXiv:0910.5670;
J.Phys. {\bf A43} (2010) 195401, arXiv:0911.2396;\\
A.Popolitov, arXiv:1001.1407;\\
Wei He and Yan-Gang Miao,
arXiv:1006.1214; arXiv:1006.5185;\\
K.Maruyoshi and M.Taki,
arXiv:1006.4505;\\
F.Fucito, J.F.Morales, R.Poghossian and D. Ricci Pacifici,
arXiv:1103.4495;\\
Y.Zenkevich,
arXiv:1103.4843;\\
N.Dorey, T.J.Hollowood and S.Lee, arXiv:1103.5726

\bibitem{DV}
R.Dijkgraaf and C.Vafa, arXiv:0909.2453

\bibitem{Awata} H.Awata and Y.Yamada, JHEP {\bf 1001} (2010) 125, arXiv:0910.4431;

\bibitem{poles}
L.Hadasz, Z.Jaskolski and P.Suchanek,
arXiv:0911.2353;
arXiv:1004.1841

\bibitem{AGTmamo}
H.Itoyama, K.Maruyoshi and T.Oota,
Prog.Theor.Phys. {\bf 123} (2010) 957-987, arXiv:0911.4244;\\
T.Eguchi and K.Maruyoshi,
arXiv:0911.4797;
arXiv:1006.0828

\bibitem{Wyl} R.Schiappa and N.Wyllard,
arXiv:0911.5337

\bibitem{MMMS} A.Mironov, A.Morozov, Sh.Shakirov,
JHEP {\bf 02} (2010) 030, arXiv:0911.5721;
Int.J.Mod.Phys. {\bf A25} (2010) 3173-3207, arXiv:1001.0563

\bibitem{FaLi} V.Fateev and A.Litvinov, JHEP {\bf 1002} (2010) 014, arXiv:0912.0504

\bibitem{ito} H.Itoyama and T.Oota,
arXiv:1003.2929

\bibitem{MManM}
A.Mironov, A.Morozov and And.Morozov,
arXiv:1003.5752

\bibitem{Awata2} H.Awata and Y.Yamada,
arXiv:1004.5122

\bibitem{Yan}
S.Yanagida,
arXiv:1005.0216

\bibitem{DFhg}
K.Maruyoshi and F.Yagi,
arXiv:1009.5553;\\
A.Mironov, A.Morozov and A.Shakirov,
arXiv:1010.1734;\\
G.Bonelli, K.Maruyoshi, A.Tanzini and F.Yagi, arXiv:1011.5417

\bibitem{PGL} A.Mironov, A.Morozov and A.Shakirov,
arXiv:1011.3481

\bibitem{surop} A.Marshakov, A.Mironov and A.Morozov,
arXiv:1011.4491

\bibitem{towaproof} A.Mironov, A.Morozov and Sh.Shakirov,
arXiv:1011.5629

\bibitem{AL}
V.Alba, V.Fateev, A.Litvinov and G.Tarnopolsky,
arXiv:1012.1312

\bibitem{Dir}
A.Mironov, A.Morozov and Sh.Shakirov, JHEP {\bf 1102} (2011) 067 arXiv:1012.3137

\bibitem{CFT} A.Belavin, A.Polyakov, A.Zamolodchikov, Nucl.Phys. {\bf B241} (1984) 333-380;\\
A.Zamolodchikov and Al.Zamolodchikov,
\emph{Conformal field theory and critical phenomena in 2d systems}, 2009 (in Russian)

\bibitem{LMNS}
G.Moore, N.Nekrasov, S.Shatashvili, Nucl.Phys. {\bf B534} (1998) 549-611, hep-th/9711108;
hep-th/9801061\\
A.Losev, N.Nekrasov and S.Shatashvili, Commun.Math.Phys. {\bf 209} (2000) 97-121, hep-th/9712241;
ibid. 77-95, hep-th/9803265

\bibitem{charex}
A.Morozov and Sh.Shakirov,
JHEP {\bf 0904} (2009) 064, arXiv:0902.2627;\\
A.Alexandrov,
arXiv:1005.5715, arXiv:1009.4887;\\
A.Morozov,
Theor.Math.Phys. {\bf 162} (2010) 1-33 (Teor.Mat.Fiz. {\bf 161} (2010) 3-40),
arXiv:0906.3518;\\
A.Balantekin, arXiv:1011.3859

\bibitem{Nek} N.Nekrasov, Adv.Theor.Math.Phys. {\bf 7} (2004) 831-864, hep-th/0206161;\\
N.Nekrasov and A.Okounkov, hep-th/0306238

\bibitem{SW} N.Seiberg and E.Witten,
Nucl.Phys., {\bf B426} (1994) 19-52, hep-th/9408099;
Nucl.Phys., {\bf B431} (1994) 484-550, hep-th/9407087

\bibitem{SWint} A.Gorsky, I.Krichever, A.Marshakov, A.Mironov, A.Morozov,
Phys.Lett., {\bf B355} (1995) 466-477, hep-th/9505035

\bibitem{5dSW} N.Nekrasov,
Nucl.Phys. {\bf B531} (1998) 323-344, arXiv:hep-th/9609219;\\
A.Gorsky, S.Gukov and A.Mironov,
Nucl.Phys., {\bf B518} (1998) 689, arXiv:hep-th/9710239;\\
A.Marshakov, A.Mironov,
Nucl.Phys., {\bf B518} (1998) 59-91,  arXiv:hep-th/9711156

\bibitem{5dB}
H.W. Braden, A.Marshakov, A.Mironov and A.Morozov, Phys.Lett.,
 {\bf B448} (1999) 195, hep-th/9812078; Nucl.Phys.,
{\bf B558} (1999) 371, hep-th/9902205

\bibitem{Kanno} H.Awata and H.Kanno,
 arXiv:0910.0083

\bibitem{MMSm3dAGT} D.Galakhov, A.Mironov, A.Morozov, A.Smirnov, arXiv:1104.2589

\bibitem{CS} E.Witten, Comm.Math.Phys. {\bf 121} (1989) 351-399; arXiv:1001.2933;
arXiv:1101.3216

\bibitem{Gukov} T.Dimofte, S.Gukov and L.Hollands, arXiv:1006.0977

\bibitem{Jap} Y.Terashima and M.Yamazaki, arXiv:1103.5748

\bibitem{DF}
Vl.Dotsenko and V.Fateev, Nucl.Phys. {\bf B240} (1984) 312-348;\\
A.Gerasimov, A.Marshakov, A.Morozov, M.Olshanetsky, S. Shatashvili,
Int.J.Mod.Phys. {\bf A5} (1990) 2495-2589;\\
A.Gerasimov, A.Marshakov and A.Morozov,
Nucl.Phys. {\bf B328} (1989) 664, Theor.Math.Phys. {\bf 83} (1990)
466-473;
Phys.Lett. {\bf B236} (1990) 269, Sov.J.Nucl.Phys. {\bf 51} (1990)
371-372

\bibitem{Rui} S.N.M.Ruijsenaars and H.Schneider,
Ann.Phys. (NY), {\bf 170} (1986) 370;\\
S.N.M.Ruijsenaars, Comm.Math.Phys., {\bf 110} (1987) 191-213;
Comm.Math.Phys., {\bf 115} (1988) 127-165

\bibitem{Kadell}
K.W.J.Kadell,
Compositio Math. {\bf 87} (1993) 5-43;
Adv.Math. {\bf 130} (1997) 33-102;
\\
J.Kaneko,
SIAM.J.Math.Anal. {\bf 24} (1993) 1086-1110

\end{thebibliography}
\end{document}